\documentclass[prd,10pt,preprintnumbers,twocolumn,numbers,sort&compress,nofootinbib,floatfix,showpacs]{revtex4}
\usepackage{amssymb}  
\usepackage{amsfonts} 
\usepackage{amsmath}
\usepackage[hypertex]{hyperref}
\usepackage{psfrag}
\usepackage{graphicx}

\newtheorem{rule-def}[theorem]{Rule}

\newcommand{\be}{\begin{equation}}
\newcommand{\eea}{\end{eqnarray}}

\begin{document}

\date{\today}

\title{Electromagnetic Mass in $(n+2)$ Dimensional Space-times}

\author{Saibal Ray\footnote {saibal@iucaa.ernet.in}}

\affiliation{Department of Physics, Barasat Government College, Kolkata 700 124, India \&
Inter-University Centre for Astronomy and Astrophysics, Post Bag 4, Ganeshkhind, Pune 411 007, India}

\begin{abstract}
 Einstein-Maxwell field equations correspoding to higher dimensional description of static spherically symmetric space-time have been solved under two specific set of conditions, viz., (i) $\rho \ne 0$, $\nu^\prime= 0$ and (ii) $\rho=0$, $ \nu^\prime\ne 0$ where $\rho$ and $\nu$ represent the mass density and metric potential. The solution sets thus obtained satisfy the criteria of being electromagnetic mass model such that the gravitational mass vanishes for the vanishing charge density $\sigma$ and also the space-time becomes flat. Physical features related to other parameters also have been discussed.
\end{abstract}

 
\maketitle

\section{Introduction}
Higher dimensional theories are of considerable interest in connection to the studies of early Universe. It is generally believed that our 4-dimensional present space-time is the compactified form of higher-dimensional manifold. This self-compactification of higher dimensions have been considered by many workers not only in the area of grand unification theory but also superstring theory \cite{sch85,wei86}. In the Kaluza-Klein type higher dimensional theories, therefore, it is a general practice to show that extra dimensions are reducible to lower one, specially in 4D, which involved in some physical processes. In the case of Kaluza-Klein type higher dimensional inflationary scenario it have been  shown by Ishihara \cite{ish84} and Gegenberg and Das \cite{geg85} that the contraction of the internal space causes the inflation of the
usual space. There are cases in FRW cosmologies where the extra dimensions contract as a result of cosmological evolution \cite{iba86}. However, in a special version of Kaluza-Klein theory with 5D mass have been considered as the fifth dimension \cite{wes83,fuk87,ban90,cha90,leo03}. In the case of Fukui \cite{fuk87} expansion of the Universe follows by the percolation of radiation into 4D space-time from the fifth dimensional mass whereas Chatterjee and Bhui \cite{cha90} show that a huge amount of entropy can be produced following shrinkage of extra-dimension which may account for the very large value of entropy per baryon observed in 4D world. Ponce de Leon \cite{leo03} has shown that the rest mass of a particle, perceived by an observer in 4D, varies as a result of the five-dimensional motion along the extra direction and in the presence of elctromagnetic field is totally of elctromagnetic origin. 

This type of models known in the literature as {\em electromagnetic mass} models, where mass is built from the electromagnetic field alone, are of considerable importance in historical as well as physical realm \cite{lor04,fey64}. Therefore, a lot of works in this direction have been carried out by several investigators \cite{flo62,coo78,tiw84,gau85,ray04} (and the references therein). In a recent publication Ray et al. \cite{ray05} have obtained a set of higher dimensional solutions corresponding to static spherically symmetric charged dust distribution. These are characterized by the fact that they represent electromagnetic mass models where gravitational mass arises from the electromagnetic field itself. They \cite{ray05} distinctly considered the following two cases: (i) $\rho \ne 0$, $\nu^\prime= 0$ and (ii) $\rho=0$, $ \nu^\prime\ne 0$ where $\rho$ and $\nu$ are, respectively, the mass density and metric potential. For both the cases, they have uniquely shown that the space-time becomes flat in the absence of charge density $\sigma$. Thus, they proved that, in general, all the higher dimensional charged dust models are of electromagnetic origin, viz., all the physical parameters originating purely from electromagnetic field. This result is equally true for the 4-dimensional case of Tiwari and Ray \cite{tiw91}.

However, it seems that both the cases also provide other simple and interesting class of solutions besides the Minkowski space-time. We have found out these hetherto unknown higher dimensional solutions corresponding to static charge fluid spheres and explored here the possibilites whether the present set of solutions also provide electromagnetic mass models or not. It has been shown that  one of such solutions corresponds to electromagnetic mass model with constant mass density and the other one leads to an $n$-dimensional Reissner-Nordstr{\"o}m metric which, in the case of perfect fluid, represents an equation of state $\rho + p = 0$ where, in general, the matter density $\rho>0$ and pressure $p<0$ \cite{tiw84}. Equation of state of this type implies that the matter distribution under consideration is in tension and hence the matter is known in the literature as a `false vacuum' or `degenerate vacuum' or `$\rho$-vacuum' \cite{dav84,blo84,hog84,kai84}.\\

\section{Basic field equations and some general results}
We start with the Einstein-Maxwell field equations for static spherically symmetric charged dust corresponding to $(n+2)$ dimensional space-time as given by Ray et al.~\cite{ray05}. The equations are 
\begin{eqnarray}
e^{-\lambda}\left[\frac{n \nu^{\prime}}{2r}+ \frac{n(n-1)}{2r^2}\right]-\frac{n(n-1)}{2r^2}= -E^{2},~~~~~~ \label{eqn:field1}\\
e^{-\lambda}\left[\frac{n \lambda^{\prime}}{2r}-\frac{n(n-1)}{2r^2}\right]+\frac{n(n-1)}{2r^2}= 8\pi\rho+E^{2}, \label{eqn:field2}\\
e^{-\lambda}\left[\frac{\nu^{{\prime}{\prime}}}{2}+\frac{{\nu^{\prime}}^{2}}{4}-\frac{
{\nu^{\prime}\lambda^{\prime}}}{4}+\frac{(n-1)(\nu^{\prime}-\lambda^{\prime})}{2r} \right. ~~~~~~~ \nonumber \\
 +\left. \frac{(n-1)(n-2)}{2r^2}\right] -\frac{(n-1)(n-2)}{2r^2}= E^{2},\\
\label{eqn:field3}
[r^{n}E]^{\prime}= 4\pi r^{n} \sigma e^{\lambda/2} ~~~~~~~~~~~~~~~~~~~~~~~~~~~~~~~~~~~
\label{eqn:field4}
\end{eqnarray}
where $\rho(r)$, $E(r)$ and $\sigma(r)$ are, respectively, the mass density, the electric field strength and charge density of the sphere of radius $r$.\\
From the field equations (\ref{eqn:field1}) and (\ref{eqn:field2}), in a straight forward way, we can have
\begin{eqnarray}
e^{-\lambda}(\nu^\prime+\lambda^\prime)= \frac{16\pi r\rho}{n}.
\label{eqn:field5}
\end{eqnarray}
Now, we make use of the conservation equations $T^i_{j;i}= 0$ which yield
\begin{eqnarray}
\rho{\nu^\prime}= \frac{1}{4\pi r^4}\frac{d}{dr}[q^2] + \frac{(n-2)E^2}{2\pi r}
\label{eqn:field6}
\end{eqnarray}
where the charge, $q$, is related with the electric field strength, $E$, through the integral form of the Maxwell's equation~(\ref{eqn:field4}), which can be written as
\begin{eqnarray}
q = Er^{n} = 4\pi\int_{0}^{r}\sigma e^{\lambda/2}r^{n}dr.
\label{eqn:field7}
\end{eqnarray}
Again, the equation~(\ref{eqn:field2}) can be expressed in the following form 
\begin{eqnarray}
e^{-\lambda}= 1- \frac{16\pi}{nr^{n-1}}\int_{0}^{r}\left(\rho+\frac{E^2}{8\pi}\right)r^{n}dr.
\label{eqn:field8}
\end{eqnarray}
If we now assume that the charge density is known, then
\begin{eqnarray}
\sigma(r)=\sigma_0 e^{-\lambda/2}
\end{eqnarray}
where $\sigma_0$ is the constant charge density at the center, $r=0$, of the spherical distribution. This, from the equation~(\ref{eqn:field7}), yield the expression for electric charge and electric field strength as
\begin{eqnarray}
q(r)= Er^{n} = \frac{4\pi \sigma_0}{n+1}r^{n+1}. 
\label{eqn:field10}
\end{eqnarray}

The gravitational mass, $m$, then can be obtained for $r>a$ by the comparison of the metric coefficients expressed in the equation~(\ref{eqn:field8}) and the following Reissner-Nordstr\"{o}m one 
\begin{eqnarray}
e^{-\lambda}= 1- \frac{2m}{r} + \frac{q^2}{r^2}
\label{eqn:field11}
\end{eqnarray}

as 
\begin{eqnarray}
m(r)= \frac{8\pi}{nr^{n-2}}\int_{0}^{r}\left(\rho+\frac{E^2}{8\pi}\right)r^{n}dr + \frac{q^2}{2r}.
\label{eqn:field12}
\end{eqnarray}
If we now consider a vanishing charge density, $\sigma$, then from the equation~(\ref{eqn:field7}) one gets for the static spherically symmetric charged dust distribution $q=0=E$. Substitution of this result in the equation~(\ref{eqn:field6}) gives the relation between the mass density and metric coefficient as
\begin{eqnarray}
\rho\nu^\prime= 0.
\end{eqnarray}

\section{New class of solutions in higher dimensions}
\subsection{The case for $\rho \ne 0$, $\nu^\prime= 0$}
With $\nu^\prime= 0$, the equation~(\ref{eqn:field5}) takes the form
\begin{eqnarray}
(e^{-\lambda})^{\prime}= -\frac{16\pi r\rho}{n}.
\end{eqnarray}
Integration with $\rho$ = constant (say, $\rho_c$) and $\lambda(0)= 0$ gives
\begin{eqnarray}
e^{-\lambda}= 1-\frac{8\pi r^2 \rho_c}{n}.
\label{eqn:field15}
\end{eqnarray}
Again, by the comparison of the relation~(\ref{eqn:field11}) with that of (\ref{eqn:field15}) we get the gravitational mass as
\begin{eqnarray}
m(r)= \frac{4\pi r^3 \rho_c}{n} + \frac{q^2}{2r}.
\label{eqn:field16}
\end{eqnarray}

It would be interesting to note here two points, first of which is, that for $n=3$, the above expression for gravitational mass corresponds to the mass given by Bonnor \cite{bon60} and Cohen and Cohen \cite{coh69}. Therefore, this simple result confirms the identification of mass as the fifth dimension by Ponce de Leon \cite{leo03}. The second one is that the constant density solution set presented here is similar to that of Bonnor \cite{bon60} in connection to the 4D charged fluid spheres. However, Bonnor \cite{bon60} raised two objections to his model regarding the limits on the radius and mass which seems valid also for our higher dimensional analogue of this Bonnor solutions. The first problem is related to the possibility that for very small radius the sphere lies well within the horizon. In the case of Lorentz-type extended electron Herrera and Varela \cite{her94} dismiss the possibility stating that in this case $q/m \approx 10^{22}$ and therefore there are no real roots for $g_{00}=0$. Regarding the second problem, where for constant $\rho$ the gravitational mass $m \rightarrow \infty $ as the radius shrinks to the center, Herrera and Varela \cite{her94} prescribe for slow and adiabatic contraction with constant $m$ and $q$, but different homogenious mass density distributions $\rho_c(a)$. However, in this connection we would like to mention here that there are, besides the solutions of Bonnor \cite{bon60} and Cohen and Cohen \cite{coh69}, some other solutions also exist in the literature where interior matter density distribution is constant in nature \cite{wil68,flo83,gro85,iva02}.

Now, from the above solution set it is evident that to be electromagnetic mass model the constant mass density $\rho_c$ should depend on the charge density $\sigma$ here. To explore this possibility let us start with the equations (\ref{eqn:field12}) and (\ref{eqn:field16}) which, after comparison and simplification, provide the following differential equation
\begin{eqnarray}
\rho_c \frac{d}{dr}(r^{n+1})= \left(\rho+\frac{E^2}{8\pi}\right)r^n .
\label{eqn:field17}
\end{eqnarray}
Therefore, after performing differentiation of the above equation~(\ref{eqn:field17}) and substitution of the value of the electric field strength from the equation~(\ref{eqn:field10}), the expression for the mass density can be obtained as
\begin{eqnarray}
\rho(r)= (n+1)\rho_c - \frac{2\pi \sigma_0^2 r^2}{(n+1)^2}. 
\end{eqnarray}
The density is regular here as the central and boundary values can, respectively, be given by 
\begin{eqnarray}
\rho_0= (n+1)\rho_c
\label{eqn:field19}
\end{eqnarray}
and 
\begin{eqnarray}
\rho_a= (n+1)\rho_c - \frac{2\pi \sigma_0^2 a^2}{(n+1)^2}. 
\label{eqn:field20}
\end{eqnarray}
It is, therefore, clear from the above two equations (\ref{eqn:field19}) and (\ref{eqn:field20}) that the mass density is decreasing from centre to boundary and likewise the case of Khadekar et al. \cite{kha01} does not vanish at the surface of the sphere. However, unless the nature of $\rho_c$ is known in terms of the charge density $\sigma$ we see that $\rho(r)$, corresponding to its central and boundary values, does not satisfy to be of electromagnetic origin. Therefore, applying boundary condition on mass density so that $\rho_a= 0$ at $r=a$ we can easily set $\rho_c = 2\pi \sigma_0^2 a^{2}/(n+1)^2$. Hence, with this value of $\rho_c$ the mass density and the total gravitational mass, respectively, become
\begin{eqnarray}
\rho(r)= \frac{2\pi \sigma_0^2}{(n+1)^2}(a^2 - r^2),
\end{eqnarray}
\begin{eqnarray}
m(a) = \frac{8\pi^2 \sigma_0^2}{n(n+1)^2}\left[na^{2n+1} + a^5\right].
\end{eqnarray}
both of which do vanish for the vanishing charge density $\sigma$. \\
However, one can also be interested, as apparently $\rho_a \ne 0$ at the boundary, to find out the effective matter-energy density which reads as
\begin{equation}
\rho_{eff} = \rho_0 - \rho_a = \frac{2\pi \sigma_0^2 a^2}{(n+1)^2}.
\end{equation}
This is exactly equal to $\rho_c$ and hence assuming $\rho_{eff}= \rho_c$ (justification of which will be given later on) one can also arrive at the same expression for the total gravitational mass. Therefore, the gravitational mass here is of purely electromagnetic origin. \\

\subsection{The case for $\rho=0$, $ \nu^\prime\ne 0$}
In this case the equation~(\ref{eqn:field5}) gives $\nu^{\prime}=-\lambda^{\prime}$ which leads to an n-dimensional Reissner-Nordstr\"{o}m metric related to the spherically symmetric static charged fluid distribution. By the use of this relation in equation (\ref{eqn:field8}) and also substituting the value of electric field strength from equation (\ref{eqn:field10}) we get
\begin{eqnarray}
e^{\nu} = e^{-\lambda}= 1- \frac{32\pi^2 \sigma_0^2}{n{(n+1)^2}(n+3)}r^4 dr.
\end{eqnarray}
Similarly, substituting the equation~(\ref{eqn:field10}) in the equation~(\ref{eqn:field12}) and setting $r=a$ for the boundary, one can get the total gravitational mass as
\begin{equation}
m(a) = \frac{8\pi^2 \sigma_0^2}{n{(n+1)^2}(n+3)}\left[n(n+3)a^{2n+1} + 2a^5\right].
\end{equation}
Obviously, the gravitational mass vanishes for the vanishing charge density $\sigma$ and also the space-times become flat. Therefore, the solution set show that the gravitational mass is originating from the electromagnetic field alone and hence corresponds to electromagnetic mass model. \\

\section{Discussions}
\begin{enumerate}
\item It is shown by Tiwari and Ray \cite{tiw84} that any four dimensional presentation of a bounded continuous static spherically symmetric charged dust solution, if exists, can only be of electromagnetic origin. This result have been extended in the higher dimensional general relativity by Ray et al. \cite{ray05} where they have studied the cases (i) $\rho \ne 0$, $\nu^\prime= 0$ and (ii) $\rho=0$, $ \nu^\prime\ne 0$. They could able to show that in the absence of charge density $\sigma$ all the physical parameters do vanish and also the space-time becomes flat. Thus it was expected that other charged dust solutions corresponding to the same cases, viz., (i) $\rho \ne 0$, $\nu^\prime= 0$ and (ii) $\rho=0$, $ \nu^\prime\ne 0$, will also provide electromagnetic mass models such that gravitational mass, including all the physical parameters, will originate from the electromagnetic field alone. However, in the present study for the above mentioned same two cases, it has been shown that the result of Tiwari and Ray \cite{tiw91} is also valid for these new type of higher dimensional solutions.

\item In the case $A$, viz., $\rho \ne 0$, $\nu^\prime= 0$ we have assumed, as a special case, that $\rho_{eff}= \rho_c$. To justify this assumption let us rewrite the expressions for the central and boundary mass densities of equations (\ref{eqn:field19}) and (\ref{eqn:field20}) in the following forms: $\rho_c= \rho_0/(n+1)$ and $\rho_c= \rho_a/(n+1) + 2\pi \sigma_0^2 a^{2}/(n+1)^3$. Therefore, we observe that $\rho_0$ and $\rho_a$ being the extream two mass densities the constant density $\rho_c$, by virtue of the above two expressions, fits in well for the effective mass density $\rho_{eff}$ which could be looked at as if falling within the range $ \rho_0 \geq \rho_{eff}= \rho_c \geq \rho_a$.

\item It is to be noted here that Tiwari et al. \cite{tiw84} exploit the relation $g_{00}g_{11} = - 1$, where $g_{00}$ and $g_{11}$ are the metric coefficients, to obtain $\rho + p = 0$ via $\lambda + \nu =0$ for the spherically symmetric Reissner-Nordstr\"{o}m solution. Therefore, by considering this metric relation to be valid inside a charged perfect-fluid distribution, it is shown by them that the mass density and the pressure of the distribution are of electromagnetic origin. However, we have arrived at the same result that in the absence of charge there exists no interior solution but in a reverse procedure in the case of higher dimensional dust case.

It is also to be noted here that for the boundary condition $\nu + \lambda = 0$ one can easily obtain $\rho + p = 0$ and vice versa, so that $\lambda = -\nu \leftrightarrow p = - \rho$ [15]. This is known as the vacuum- or imperfect-fluid equation of state in the literature \cite{dav84,blo84,hog84,kai84}. We would like to mention here that the equation of state in the form $p + \rho = 0$ is discussed by Gliner \cite{gli66} in his study of the algebraic properties of the energy-momentum tensor of ordinary matter through the metric tensors.
\end{enumerate}

\section*{Acknowledgment}
The author's thanks are due to the authority of IUCAA, Pune for providing him Associateship programme under which a part of this work was carried out. Support under UGC grant (No. F-PSN-002/04-05/ERO) is also gratefully acknowledged.

{}
\end{document}